# Effects of Epileptiform Activity on Discharge Outcome in Critically Ill Patients: A Retrospective Cross-Sectional Study

Harsh Parikh, MS*[1], Kentaro Hoffman, PhD*[2], Haoqi Sun, PhD*[3], Sahar F. Zafar, MD *[3], Wendong Ge, PhD[3], Jin Jing, PhD[3], Lin Liu, PhD[4, +], Jimeng Sun, PhD[#5], Aaron Struck, MD[6], Alexander Volfovsky, PhD**[1], Cynthia Rudin, PhD**[#1], M. Brandon Westover, MD, PhD**[#3]

[1]Duke University, Dept. of Computer Science

[2]University of North Carolina at Chapel Hill, Dept. of Statistics and Operation Research

[3]Massachusetts General Hospital, Dept. of Neurology

[4]Harvard T. H. Chan School of Public Health, Dept. of Epidemiology

[5]University of Illinois Urbana-Champaign, The Grainger College of Engineering

[6]University of Wisconsin-Madison Dept. of Neurology

[+]Present address: Institute of Natural Sciences, MOELSC, School of Mathematical Sciences and SJTU-Yale Joint Center for Biostatistics and Data Science, Shanghai Jiao Tong University, Shanghai, China

*Co-first authors    **Co-senior authors    # Full professors

Corresponding author:

M. Brandon Westover, M.D. Ph.D.

Massachusetts General Hospital

55 Fruit Street, Boston, MA 02114

mwestover@mgh.harvard.edu

**Research in context**

**Evidence before this study**
Several prior studies have established associations between epileptiform activity (EA) and neurologic outcomes. However, these studies have not adjusted for the treatment effects of anti-seizure medications (ASM). Not adjusting for treatment is problematic because several recent studies suggest aggressive ASM use, especially intravenous anesthetic drugs like propofol, may be harmful. However, adjusting for these factors is challenging because of the complex interactions and feedback between EA and ASM. Yet, without adjusting for these factors, it remains unclear whether associations between EA and poor outcomes are due to over-treatment, underlying illness, or effects of EA. Whether (aggressive) treatment is needed has been a subject of debate in the field without a definitive answer.

**Added value of this study**
Using a novel causal inference approach and neurologist validation, our results provide a more accurate estimate of the effect of EA on neurologic outcome based on a large retrospective cross-sectional dataset.

**Implications of all the available evidence**
The result favors treating since EA burden indeed worsens neurologic outcomes after adjusting for ASMs. However, the effect depends on the pattern of EA (maximum and average EA burden). An optimal treatment policy that reduces the EA burden is needed to improve patient outcomes.

**Summary**

**Background** Epileptiform activity (EA) is associated with worse outcomes including increased risk of disability and death. However, the effect of EA on neurologic outcome is confounded by the feedback between treatment with anti-seizure medications (ASM) and EA burden. A randomized clinical trial is challenging due to the sequential nature of EA-ASM feedback, as well as ethical reasons. However, some mechanistic knowledge is available, e.g., how drugs are absorbed. This knowledge together with observational data could provide a more accurate effect estimate using causal inference.

**Methods** We performed a retrospective cross-sectional study with 995 patients with the modified Rankin Scale (mRS) at discharge as the outcome and the EA burden defined as mean or maximum proportion of time spent with EA in six-hour windows in the first 24 hours of electroencephalography as the exposure. We estimated the change in discharge mRS if everyone in the dataset had experienced a certain EA burden and were untreated. We combined pharmacological modeling with an interpretable matching method to account for confounding and EA-ASM feedback. Our matched groups' quality was validated by the neurologists.

**Findings** Having maximum EA burden ≥ 75% when untreated had a 22.2% increased chance of a poor outcome (severe disability or death); mild but long-lasting EA (average EA burden ≥ 10%) increased the risk of a poor outcome by 13·52%. The effect sizes were heterogeneous depending on pre-admission profile, e.g., patients with hypoxic ischemic encephalopathy (HIE) or acquired brain injury (ABI) were more affected.

**Interpretation** Interventions should put a higher priority on patients with an average EA burden higher than 10%, while treatment should be more conservative when maximum EA burden is low. Treatment should also be tailored to individual pre-admission profile because the potential for EA to cause harm depends on age, past medical history, and reason for admission.

**Funding** NIH (R01NS102190, R01NS102574, R01NS107291, RF1AG064312, RF1NS120947, R01AG073410 K23NS114201), NSF (IIS-214706, DMS-2046880).

**Keywords**

Epileptiform activity; modified Rankin Scale; anti-seizure medications; causal inference

**Introduction**

Epileptiform activity (EA, also referred as ictal-interictal-injury continuum, IIIC[1] activity) is common in critically ill patients, affecting more than half of patients who undergo electroencephalography (EEG) in critical care[2–5]. EA varies in terms of spatial extent (generalized vs. lateralized) and periodicity (periodic vs. rhythmic vs. sporadic). Here, we consider EA to be the combination seizures, lateralized periodic discharges (LPD), generalized periodic discharges (GPD), and lateralized rhythmic delta activity (LRDA). Prolonged EA is associated with in-hospital mortality, and survivors often suffer long-term functional and cognitive disability[6–9]. Despite a growing literature indicating EA is associated with poor outcomes[10], there is a long-standing debate about whether EA is part of a causal pathway that worsens outcomes and thus requires aggressive treatment, or worsened outcomes are due to mechanisms other than EA such as medication side effects or the inciting illness, with EA as an epi-phenomenon[11–16].

Studies of the effects of EA on neurologic outcomes to date have suffered from a variety of limitations. First, a hypothetical clinical trial studying the effect of untreated EA would compare outcomes in groups of patients who experience different levels of EA burden but are otherwise matched for relevant clinical variables, while ensuring no ASMs are administered, which is neither possible nor ethical. Second, observational data contain complex interactions of EA and anti-seizure medications (ASM), i.e., physicians administer ASMs based on patients' EA, and in turn, EA is affected by ASMs. This creates entanglement (Figure 1) between EA and ASMs, obscuring the true effect of EAs. Further, observational datasets are subject to unmeasured confounding. Therefore, a naïve statistical analysis can have higher bias and variance. Prior studies of EA have used regression models to adjust for medical history and demographic factors[7–9,17] and interpreted the regression coefficient for EA as the effect of EA on the outcome. While this approach is appealing for its simplicity, the interpretation of regression coefficients can be misleading due to EA-ASM interactions. On the other hand, relying on data-driven black-box machine learning models for such analyses lead to uninterpretable conclusions and difficult clinical validation.

Here, our objective is to quantify the heterogeneous effects of EA with an interpretability-centered approach where a physician can verify the quality of every analysis step, including how a current patient compares to others (case-based reasoning), how drug absorption and response is modeled, and the relative importance of covariates. We leveraged the domain knowledge using pharmacokinetic-pharmacodynamic (PK/PD) models to describe the interactions between clinical decisions and physiological response, which identifies individuals who react similarly to treatments. We used a matching method to estimate both medium and long-term effects of clinical decisions and physiological responses. The matched group constructed for each patient can be validated via chart review.

**Methods**

*Study design*

We performed a retrospective cross-sectional study of ICU patients admitted to the Massachusetts General Hospital (MGH) between December 1st, 2011 to October 14th, 2017. Institutional review boards at MGH, Duke University, and University of North Carolina at Chapel Hill approved the analysis without requiring written informed consent. The inclusion criteria were (1) a clinical neurophysiologist or epileptologist read the reports of EEG findings in the electronic health record of MGH and identified electrographic EA; and (2) at least 18 years old. The exclusion criteria (Figure S1) are (1) low EEG quality, where the duration of consecutive artifact (defined in "EEG pre-processing and artifact detection" in Supplementary Material) is more than 30% of the total length; (2) less than 2 hours of continuous EEG monitoring; and (3) missing outcome or covariates. The results are reported in accordance with the Strengthening the Reporting of Observational Studies in Epidemiology (STROBE) guidelines for reporting observational studies[18].

*Outcome: discharge mRS*

The outcome is the modified Rankin Scale (mRS) at hospital discharge. The mRS is a 0 to 6 ordinal scale where 0 means no symptoms and 6 means dead. We dichotomized mRS into poor (mRS ≥ 4: moderately severe disability) and favorable (mRS ≤ 3: moderate disability) outcomes[9]. Patients with missing discharge mRS (n = 8) were excluded. The outcome adjudicators were blinded to the EA status and burden.

*Exposure: EA burden*

EA is defined as one of four patterns[1]: (1) generalized periodic discharges (GPD), (2) lateralized periodic discharges (LPD), (3) lateralized rhythmic delta activity (LRDA) and (4) Seizure (Sz). Every two-second EEG segment was classified as containing EA or not by a deep neural network (Figure S5 and Figure S6) using an automated algorithm that was developed to detect these key ICU EEG patterns (rather than relying on EEG reports). Then a timeseries was generated as the fraction of two-second EEG segments containing EA over a six-hour window. We chose six hours to observe the effects of ASMs on EA and for physicians to adjust ASM treatment. EA burden is defined in two clinically meaningful ways: (1) $E_{mean}$: measures the average EA fraction among all six-hour sliding windows (step size of ten minutes) within the first 24 hours of EEG; and (2) $E_{max}$: measures the maximum EA fraction among all six-hour sliding windows within the first 24 hours of EEG. By quantifying EA burden in these two ways, we sought to separate potentially different effects of intense but brief EA ($E_{max}$) from prolonged periods of less intense EA ($E_{mean}$). For interpretability and statistical efficiency, we binned $E_{max}$ burden into following four levels – mild (0% to 25%), moderate (25% to 50%), severe (50% to 75%), very severe (75% to 100%) and $E_{mean}$ into following four levels - mild (0% to 2%), moderate (2% to 10%), severe (10% to 30%), very severe (30% to 100%). Like for EA, we binned each administered ASM into two groups: (a) minimal or low and (b) significant. Bins were chosen as a function of the estimated median ED50 across the patient population. Here, for a particular ASM, a patient's treatment for that ASM is considered "minimal" if the mean dose of that ASM for the patient is less than 1/10th of the population median ED50, otherwise it is considered "significant". If all ASMs administered for each patient were in the "minimal" category then we characterize that patient's overall ASM regime as "untreated", otherwise if any ASM administration was "significant" then the patient is considered "treated" with ASMs. A sensitivity analysis of these choices is provided in "Sensitivity analysis" in Supplementary Material.

*Covariates: pre-admission variables and drug responsiveness*

First, for each patient we observed 70 covariates covering demographics (age, sex, and marital status), clinical factors (history of seizures or epilepsy, chronic kidney disease, etc.), and admission diagnosis (cancer, subarachnoid hemorrhage, etc.). They are denoted as the pre-admission variables. The primary ones are listed in Table 1 and the full list in Table S1. Patients with missing covariates (n = 314) were excluded (Figure S1).

The second source of confounding comes from a patient's drug responsiveness. Due to differing medical history, medical conditions, age, etc., patients may respond differently to ASM. The ASM studied here are lacosamide, levetiracetam, midazolam, pentobarbital, phenobarbital, propofol, valproate, lorazepam, diazepam, and combined phenytoin and fosphenytoin. While other ASMs are sometimes used and can be effective in treating EA, their usage was much less frequent in our cohort. To account for this, we model each patient's response to ASM via one-compartment pharmacokinetics (PK) models and the Hill equation for pharmacodynamic (PD) response. For the PK model, the half-lives of ASM were obtained from drug databases and fixed (Table S2). For the PD model, the Hill coefficient and the dosage required to reduce EA burden by 50% ($ED_{50}$) represent the patient's drug responsiveness for each ASM and are estimated from the data. Estimation of per-patient PD models allowed adjusting for heterogeneity of drug-response and drug-effectiveness (i.e., the effect on EA burden).

Note that, we do not explicitly model drug interactions to avoid curse of dimensionality issues. Nevertheless, most of the ASMs used in our cohort are levetiracetam, lacosamide, and propofol for which interactions are minimal, and ASMs known to interact are rare in our study. As our study focuses on the time that patients are on EEG (average duration 27.25 hours), PKPD changes are likely to be small.

Patients with missing EEG or ASM data (n = 471) were not included according to the inclusion criteria. However, we found no significant difference between the median of discharge mRS (outcome) and those without missing EEG data (p = 0.39), as detailed in "Missingness Pattern" in the Supplementary Material.

*Effect estimation through matching*

We aim to estimate the degree to which untreated epileptiform activity worsens neurological outcomes. Our estimand is the probability of a poor outcome if the patient has EA burden ($E_{max}$ or $E_{mean}$) equal to a given level in the absence of treatment. We study this counterfactual outcome (what would have happened without ASM) because it disentangles the effects of EA from ASM on outcomes.

The covariates, including age, demographics, clinical factors such as disease histories and medical diagnoses, and drug response parameters, are used for matching; each patient is matched to patients with similar covariates. We use an interpretable distance metric-based matching algorithm, Matching After Learning To Stretch (MALTS)[19], to match patients directly on covariates (not on proxies like propensity scores). The resulting matched groups permit case-based reasoning and allow estimating the heterogeneous effects of both EA and drugs on outcomes. The overall analysis framework is shown in Figure 4.

*Neurologist review of the matched groups*

Inspired by similar approaches in the social sciences[20], one can check for unobserved confounders by having a domain expert perform a post-facto analysis of matched groups. Three independent neurologists (MBW, SZ, AS) were sent three randomly chosen matched groups for manual chart review to assess matching quality and determine unmeasured confounding. Reviewers were asked to independently perform a qualitative analysis of the matched groups and report the estimated chance of experiencing a high EA burden and the chance of a poor outcome, according to bins of 0-20%, 20-40%, 40-60%, 60-80%, and 80-100%. A successful validation means that: (1) in the clinician's judgement, the patients in the matched groups are medically similar, i.e., the matched group is tight; and (2) the outcome prognosis and EA propensity are similar. Similarly, a failed validation means that the reviewing clinicians find a significant medical difference in either outcome prognosis or EA propensity that, in the clinicians' judgement, would invalidate the matching group. The tightness of the matched groups is important for validation. The tighter the matched groups, the more our study resembles a randomized control trial with matched treatment and control patients.

*Statistical analysis*

Continuous data is presented using the median and interquartile range. Categorical data is presented using number and percentage. Confidence intervals are derived via bootstrapping, by randomly sampling the same number of patients with replacement 1000 times and computing lower and upper bounds as the 2.5% and 97.5% percentile of the bootstrapped results respectively.

Sensitivity analysis was conducted in terms of unobserved confounding and imprecision in EA burden, by varying the strength of unobserved confounding and noise in the EA burden respectively, as detailed in the "Sensitivity analysis" in Supplementary Material.

*Role of the funding source*

The funding source was not involved in any part of research.

# Results

*Average effect of maximum EA burden on outcome*

Figure 2A illustrates the first main result: patients with higher levels of $E_{max}$ are at higher risk of poor neurologic outcomes. Moreover, the risk of a poor outcome increases monotonically as EA burden increases, culminating in an average increase of 22.3% when a patient's untreated EA burden increases from mild (0 to 25%) to very severe (75% to 100%).

*Average effect of mean EA burden on outcome*

Figure 2B shows the other main result: those with higher $E_{mean}$ are also at higher risk of a poor outcome. Similar to $E_{max}$, the risk increases monotonically with increase in EA burden. Our results indicate that severe and very severe prolonged EA burden (over 24 hours) increase the risk of a poor outcome by 18.0% compared to mild prolonged EA burden.

*Heterogeneity in effects for maximum EA burden*

The features to search for interesting subgroups were chosen a priori based on patients' history and diagnosis. Figure 3 breaks down the population into subpopulations based on various central nervous system (CNS) and medical conditions with differing conditional average treatment effects. Our results indicate that patients with hypoxic-ischemic encephalopathy (HIE) or acquired brain injury (ABI) are at higher risk of a worse outcome in response to a large $E_{max}$, possibly attributable to inflammation leading to exacerbated harm of neurologic injury. We also examined race and sex as possible effect modifiers of EA burden, which did not modify the risk from $E_{max}$. These results suggest (though in most cases not statistically significant, given the size of the subgroups) potential effect modification by many different types of pathology. Our results suggests that future, larger sample studies should be designed to better understand the heterogeneity of effects of EA for patients with different CNS pathologies.

*Interpretation of matched group*

MALTS provide the relative importance of covariates used in matching. As shown in Figure S3, for $E_{max}$, two measures of illness severity were heavily weighted (worst GCS in 1st 24 hours and APACHE II). Further, levetiracetam pharmacodynamics (ED50 and Hill Coefficient) and diastolic blood pressure were the other most important variables. These observations suggest that our matched groups consist of individuals that agree on factors representing overall health, current level of neurologic impairment and their responsiveness to important NSAED. In Figure S3, one can see that the three least important matching variables are Hill coefficients and $ED_{50}$ parameters from one of the anti-seizure medications. This stands in contrast with $ED_{50}$ for levetiracetam, one of the top five most important variables. This suggests that information about responsiveness to levetiracetam, a potent non-sedating anti-seizure drug is more critical in estimating effects of EA on outcome than other less potent ASMs.

*Matched groups are validated by neurologists' chart review*

The validity of the effect estimation depends on the validity of the matching groups. Our physicians evaluated a random subset of 3 matched groups using clinical notes (hospital admission summaries). Note that these clinical notes (and the information in them) were not explicitly used by our method for matching; they were only used as an independent method for evaluating the validity of the matched groups. This allowed the physician reviewers to reason about the existence of any important unobserved confounders that could invalidate the analysis.

As shown in Table 2, the neurologists found no problematic sources of confounding. Moreover, we can observe which factors each group was matched tightly on. For example, group #2 is tightly matched with patients having similar initial Glasgow Comma (iGCS) and APACHE II scores and all but one having relatively good prognoses. By contrast, group #3 is tightly matched on acute neurological injuries at the cost of a looser match on APACHE II scores. Viewing what is tightly matched in each group provides a holistic evaluation of the factors controlled for, e.g., age, and the factors either unimportant or with small sample size, e.g., many of the less common medical conditions.

**Discussion**

*Clinical Implications.* Our findings have two primary implications for treatment of EA. (1) Treatment should be based on both $E_{max}$ and $E_{mean}$. Intense bursts of EA burden (captured by $E_{max}$), even if relatively brief (6 hours) lead to worse outcomes. Similarly, sustained periods of EA (captured by $E_{mean}$) show a monotonic relationship with the outcome: EA < 2% has minimal effect, but any EA ≥ 10% increases the risk of a worse outcome at least by 13.5%. This suggests interventions should put a higher priority on patients with a mean EA burden higher than 10%, while treatment should be more conservative when maximum EA burden is low. (2) Treatment should also be tailored to individual pre-admission profile because the potential for EA to cause harm depends on age, past medical history, and reason for admission. By contrast, current treatment protocols tend to be generic, based on the duration of EA but providing little guidance on how to consider other patient characteristics. As a result, treatment approaches vary widely between doctors.

*Results in context.* Our work builds on prior results demonstrating associations between EA, treatments, and neurologic outcomes. Oddo et al[10] studied 201 ICU patients of which 60% had sepsis as an admission diagnosis. They found that EA (seizures and periodic discharges) was associated with worse outcomes based on a regression adjustment for age, coma, circulatory shock, acute renal failure, and acute hepatic failure. However, they did not adjust for treatment with ASM, including phenytoin (given to 67% of patients), levetiracetam (62% of patients), and lorazepam (57% of patients), and four other drugs. Tabaeizadeh et al.[21] found that the maximum daily burden of EA/seizures is associated with a higher risk of poor outcomes in 143 patients with acute ischemic stroke. However, they did not control for ASMs which were given to 83% of patients. Lack of adjusting for drug use is also found in the pediatric literature on EA[6]. Not adjusting for treatment is problematic because a growing number of studies suggest aggressive ASM use, especially with intravenous anesthetic drugs like propofol, may be harmful. For example, a recent retrospective study by Marchi et al.[22] of 467 patients with incident status epilepticus found that therapeutic coma was associated with poorer outcome, higher prevalence of infection, and longer hospital stay[23,24]. However, because more aggressive treatment is reserved for more severely ill patients, these studies have come under criticism for failing to adequately adjust for the type and severity of medical illness, and for the burden of epileptiform activity.

*Generalizability.* The monotonic relationship between EA burden and poor outcome has been suggested by multiple studies[6]. Our study supports these findings and suggests high likelihood of causal relationship. We expect a causal relationship to be generalizable, particularly in light of prior studies that have found similar relation between EA burden and outcomes in disease subgroups included in our study e.g., stroke, cardiac arrest, non-neurological medical and surgical patients and in distinct populations not included in our work (e.g., pediatric ICU patients[6]). Although randomized trials are currently challenging to conduct, alternative epidemiological approaches could be used to test the results in an independent dataset.

*Methodological improvement.* A key component of our approach is adjusting for patients' drug responsiveness (PK and PD parameters) to account for patient heterogeneity. Critically ill patients can be different in many ways including measured and unmeasured variables. By accounting for individual drug responsiveness, we were able to adjust for exposure to anti-seizure drugs, such as phenytoin and pentobarbital, where the medications themselves may worsen outcomes. Another advance is our application of a methodology designed specifically for causal inference using observational data. The matching approach in MALTS achieves both the flexibility of being free of model misspecification (non-parametric) and the interpretability of the estimated weights, creating less biased effect estimates. With this new approach, we provide credible estimates of how much harm EA causes and in which types of patients.

*Limitations.* Our approach has several limitations. Although we have conducted sensitivity analysis, this does not change the retrospective cross-sectional nature of the study. Therefore, there are still unmeasured confounding, such as illness present before admission not captured in our data that positively contributes to both having high EA burden and poor outcome. Delirium can be a cause of lower GCS, however, it was not routinely assessed in the ICUs of our cohort. When evaluating EA burden, we did not consider the subtype of EA (Sz/GPD/LPD/LRDA), discharge frequency, and the spatial extent of EAs. Further, as BIPDs are rare, the raters in our study labelled BIPDs as GPDs. Our PK/PD model might be able to be further improved by including more mechanistic or physiological detail, such as a learning heterogeneous pharmacokinetic parameters across patients, adjusting for changes in PKPD parameters over the course of treatment, drug interactions and context-sensitive half-life for propofol[25]. In addition, the deep learning model that detects EA patterns is not perfect (see Figure S6), even though the result is robust to systematic shift in the model output probability (Figure S11). The current study also did not estimate an optimal treatment policy to improve patient outcomes, which is an important future research direction. We hope to organize an international data-sharing collaboration in the near future to make further progress on this important topic. An ideal observational cohort will have at least three important properties: (1) Large size: a larger dataset would allow to test for effect heterogeneity across various etiopathogenesis. (2) Multicenter design: allowing for greater variation in practice patterns. (3) Additional information: detailed annotation based on EA subtypes, potential confounders, and effect modifiers over the course of treatment.

## Contributors

HP, KH, HS, AV, CR, MBW conceptualized the study. WG, JJ, SZ, MBW curated the data. WG and JJ developed the neural network model for detecting EA in raw EEG data. HP, KH, HS analyzed the data. MBW and JS acquired funding. HP, KH, HS, AV, CR, MBW created the methodology. AV, CR, MBW gave supervision. AS, SZ, MBW performed chart review on the matched groups. HS verified the data. HP, KH, HS, created the figures. HP, KH, HS, AV, CR, MBW wrote the original draft. All authors reviewed and edited the final manuscript. HS, SZ, AV, CR, MBW had access to raw data and final responsibility for the decision to submit for publication.

## Declaration of interests

MBW is a co-founder of Beacon Biosignals, which played no role in this study. SZ is clinical neurophysiologist consultant for Corticare. All other co-authors report no competing interests.

## Funding source

Dr. Westover is supported by NIH grants: R01NS102190, R01NS102574, R01NS107291, RF1AG064312, RF1NS120947, and R01AG073410. Dr. Rudin, Dr. Volfovsky and Mr. Parikh are supported by NSF IIS-2147061. Dr. Volfovsky is also supported by NSF DMS-2046880.

## Data sharing

Written requests for access to the data reported in this paper will be considered by the corresponding author and a decision made about the appropriateness of the use of the data. If the use is appropriate, a data-sharing agreement will be put in place before a fully de-identified version of the dataset used for the analysis with individual patient data is made available.

# Figure Legends

**Observational data: What Happened**

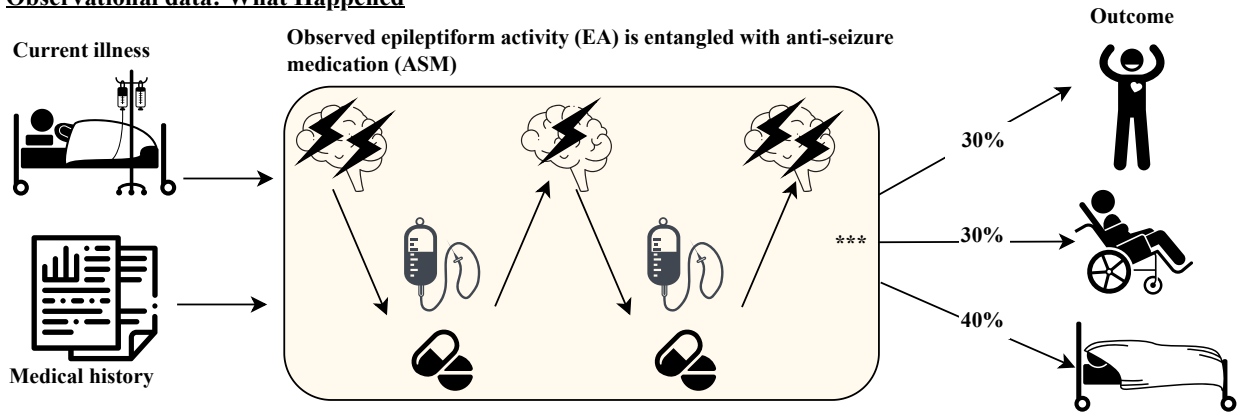

**Counterfactual: What would happen if the patient experienced different level of EA**

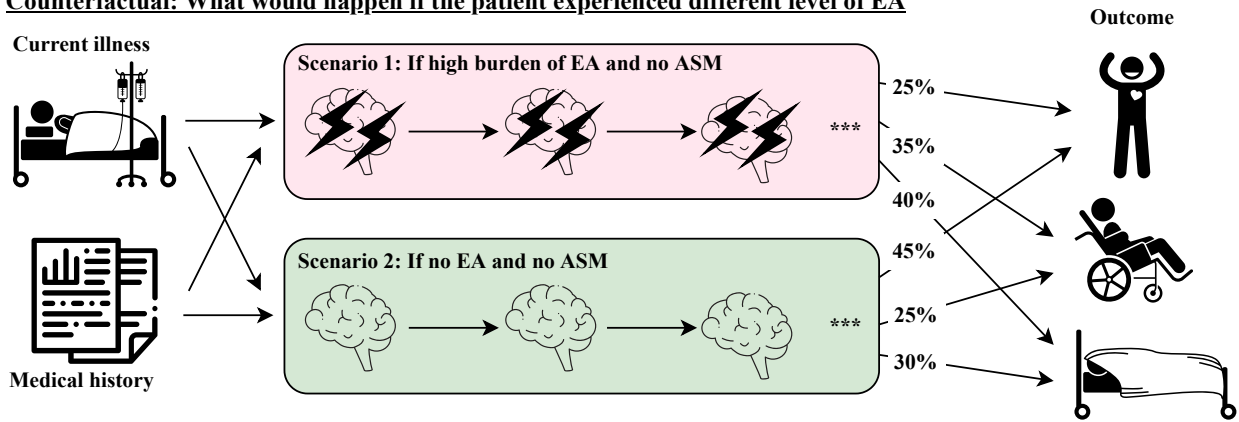

**Figure 1. Illustration of the scenarios.** *Upper:* Illustration showing that observed epileptiform activity (EA) forms a feedback loop with treatment decision, that is also influenced by current illness and medical history (left side). The entire time-series of EA and ASM influence patient outcomes. Possible outcomes include return to normal health, disability, or death at the time of hospital discharge (right side). *Lower:* Our goal is to estimate the counterfactual case: if the patient experienced different level of EA effect

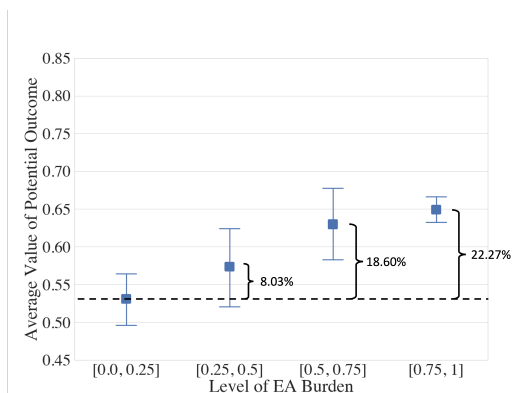

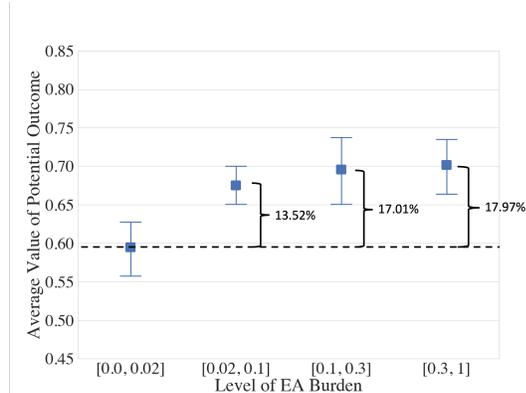

(A)                                    (B)

**Figure 2. The probability of a poor outcome mRS for Mild, Moderate, Severe, or Very Severe EA burdens.** EA Burden is quantified as (A): $E_{max}$ and (B) $E_{mean}$. In both scenarios, an increase in EA burden leads to a worse potential outcome. Outcome worsens monotonically for $E_{max}$, whereas for $E_{mean}$, there is a jump at approximately 0·5. In both plots, the horizontal line represents the baseline median average potential outcome for the mild case.

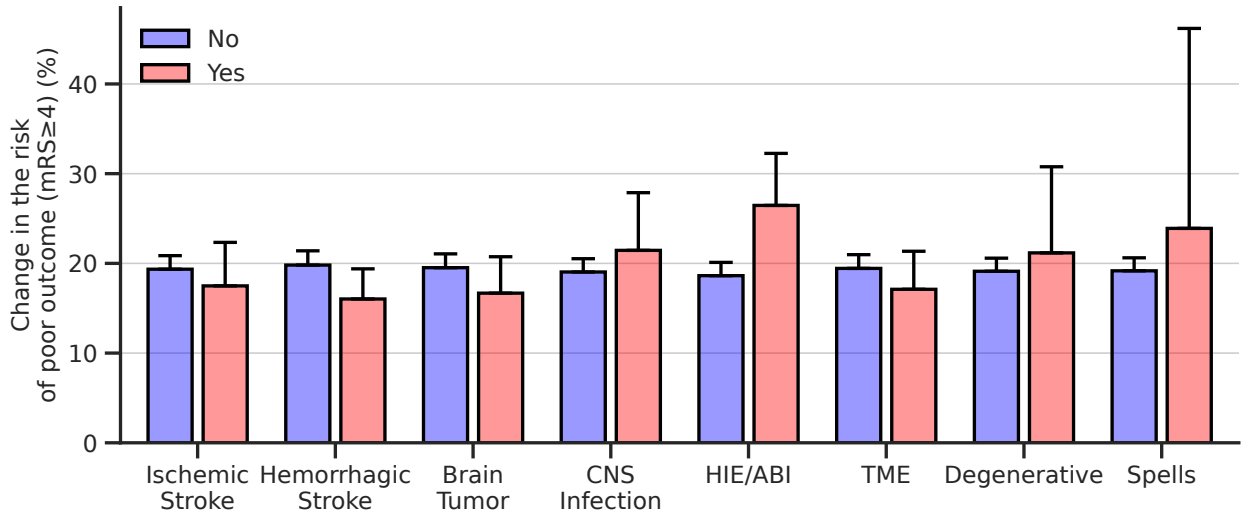

**Figure 3. Heterogeneity in the effects of $EA_{max}$.** The causal effect of EA are stratified by various central nervous system pathologies. The blue bars represent causal effects when a particular pathology is absent and the red bards represent conditional causal effects when that pathology is present.

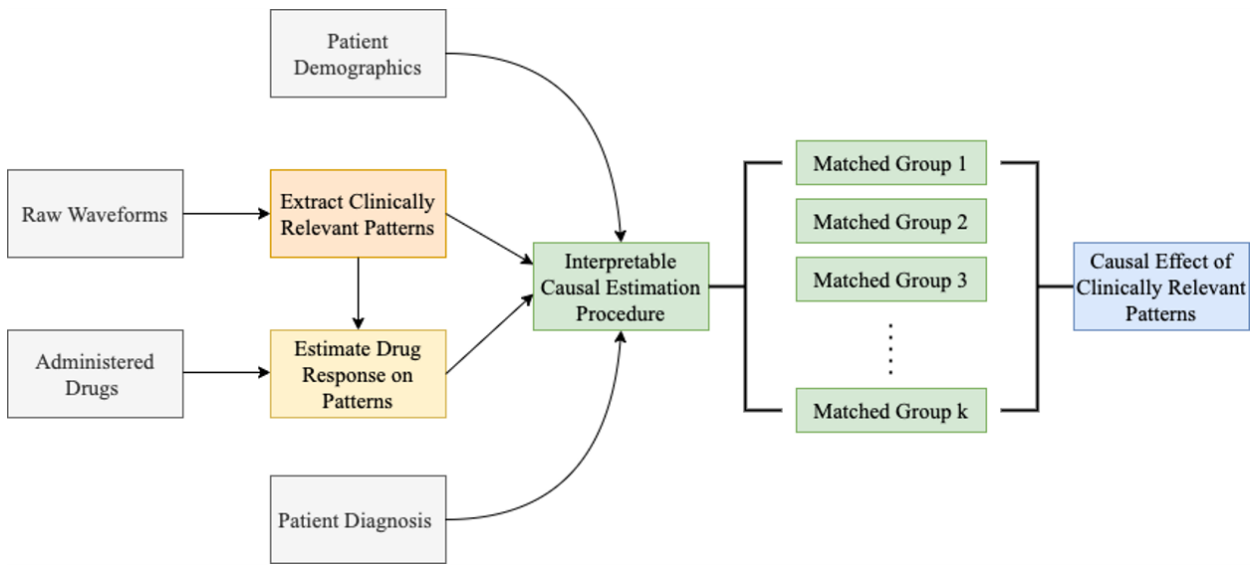

**Figure 4. The overall analysis framework.** Analysis framework consists of four parts (indicated by different colors): EA burden computation, individual PK/PD modeling, MALTS matching, and effect estimation.

Table 1. Patient characteristics

| Variable | Value |
|---|---|
| **Major covariates included as confounders (others in Table S1)** | |
| Age, year, median (IQR) | 61 (48 -- 73) |
| Sex, number of male (%) | 475 (48%) |
| Race | |
|   Asian, n (%) | 33 (3%) |
|   Black / African American, n (%) | 72 (7%) |
|   White / Caucasian, n (%) | 751 (75%) |
|   Other, n (%) | 50 (5%) |
|   Unavailable / Declined, n (%) | 84 (8%) |
| Premorbid mRS before admission, median (IQR) | 0 (0 -- 3) |
| APACHE II in first 24h, median (IQR) | 19 (11 -- 25) |
| Initial GCS, median (IQR) | 11 (6 -- 15) |
| **Exposure** | |
| EA burden: $EA_{max}$, median (IQR) | 0·65 (0·16 -- 0·99) |
| EA burden: $EA_{mean}$, median (IQR) | 0·09 (0·014 -- 0·31) |
| **Outcome** | |
| Discharge mRS, median (IQR) | 4 (4 -- 5) |

**Table 2. Three randomly chosen matched groups.** We include doctors' prognosis and their qualitative estimate of risk of high EA burden. The notes column presents neurologists' remarks during chart review, based on medical notes not used for matching. The last row contains neurologists' notes on quality of matched groups.

| Gender | Age | iGCS-Total | APACHE II (1st 24h) | Neurologists' estimates of High EA | Neurologists' estimates of Good Prognosis | Summary |
|---|---|---|---|---|---|---|
| **Matched Group 1** Diameter: 41.21 | | | | | | |
| 1 | 57 | 13 | 12 | 80-100 | 40-60 | 57M h/o LTx, SDH, epilepsy p/w seizures, PNA, and confusion. |
| | | | | 40-60 | 40-60 | |
| | | | | 60-80 | 40-60 | |
| 0 | 51 | 12 | 7 | 40-60 | 40-60 | 51F s/p renal+pancreas tx, recent L clinoid meningioma resection, p/w lethargy, confusion -> CT = L cerebral convexity collection. |
| | | | | 20-40 | 40-60 | |
| | | | | 30-50 | 50-70 | |
| 0 | 47 | 15 | 4 | 80-100 | 80-100 | 47F w/ refractory epilepsy p/w increased frequency of complex partial seizures. |
| | | | | 80-100 | 80-100 | |
| | | | | 80-100 | 80-100 | |
| 1 | 56 | 15 | 4 | 40-60 | 60-80 | 55 p/w L facial droop + aphasia. MRI negative for stroke. |
| | | | | 0-20 | 80-100 | |
| | | | | 10-20 | 50-70 | |
| 0 | 58 | 15 | 8 | 20-40 | 60-80 | 58F w/ szs 2/2 L motor strip meningioma, p/w RLE weakness -> admitted for resection |
| | | | | 0-20 | 60-80 | |
| | | | | 5-25 | 50-70 | |
| 0 | 65 | 15 | 5 | 80-100 | 40-60 | 65F w/ GBM p/w increased L weakness and staring episodes over 2 months |
| | | | | 40-60 | 40-60 | |
| | | | | 40-50 | 40-50 | |
| 1 | 29 | 6 | 16 | 20-40 | 40-60 | 29M hit by car, found with GCS6, CT = L left frontal/temporal SAH + ?midbrain ICH |
| | | | | 20-40 | 20-40 | |
| | | | | 5-25 | 30-50 | |
| 0 | 42 | 15 | 2 | 40-60 | 60-80 | 41F w/ GBM c/b seizures, resection 4 mo ago, p/w surgical wound infx, undwent I&D. |
| | | | | 0-20 | 60-80 | |
| | | | | 10-20 | 40-60 | |
| 0 | 45 | 13 | 9 | 20-40 | 40-60 | 45F p/w HA x 7d, CTA = ruptured LMCA aneurysm |
| | | | | 20-40 | 60-80 | |
| | | | | 10-20 | 40-60 | |
| 1 | 61 | 15 | 3 | 40-60 | 40-60 | 61M w/ metastatic NSCLC, p/w confusion + new brain mass. ?mets to brain. |
| | | | | 20-40 | 40-60 | |
| | | | | 5-25 | 30-50 | |
| 1 | 42 | 15 | 3 | 20-40 | 40-60 | 42M p/w WHOL -> F3, HH2 SAH from 6mm anterior communicating aneurysm rupture. |
| | | | | 20-40 | 60-80 | |
| | | | | 10-30 | 40-60 | |
| 0 | 46 | 15 | 6 | 20-40 | 40-60 | 46F p/w WHOL -> HH2, F3 SAH (right supraclinoid ICA aneurysm) -> MGH |
| | | | | 20-40 | 60-80 | |
| | | | | 10-20 | 40-60 | |

**Neurologists' notes:** The units in this matched group are similarly aged except one 29y old (involved in a car accident). The same individual also has lower GCS compared to rest of the group. In general, the 3mo prognosis of patients in the group is similar and moderately good. However, the patients have varied underlying risk of EA burden.

**Matched Group 2**
Diameter: 45.20

| | | | | | | |
|---|---|---|---|---|---|---|
| 0 | 47 | 15 | 4 | 80-100 | 80-100 | 47F w/ refractory epilepsy p/w |
| | | | | 80-100 | 80-100 | increased frequency of complex partial seizures. |
| | | | | 80-100 | 80-100 | |
| 1 | 20 | 15 | 1 | 0-20 | 80-100 | 19M h/o psychosis, 15q3 micordeletion, p/w increased L-sided tone + mutism, ?catatonia. |
| | | | | 0-20 | 80-100 | |
| | | | | 0-20 | 70-90 | |
| 0 | 30 | 15 | 3 | 20-40 | 80-100 | 29F w/ gastic bypass surgery & ?PNES, p/w acute L sided weakness, MRI negative -> PNES vs sz? |
| | | | | 0-20 | 80-100 | |
| | | | | 10-30 | 70-90 | |
| 0 | 65 | 15 | 5 | 80-100 | 60-80 | 65F w/ GBM p/w increased L weakness and staring episodes over 2 months |
| | | | | 60-80 | 60-80 | |
| | | | | 40-50 | 40-50 | |
| 0 | 31 | 12 | 15 | 0-20 | 60-80 | 30F p/w 2 wks sore throat, HA, BLE weakness, sensory level T2, spine MRI = T9 hyperintensity. |
| | | | | 0-20 | 60-80 | |
| | | | | 10-30 | 30-50 | |
| 1 | 28 | 14 | 1 | 0-20 | 80-100 | 27M w/ epilepsy vs PNES, p/w episodes of UE tonic stiffening. Admit from ED for LTM. |
| | | | | 0-20 | 80-100 | |
| | | | | 10-30 | 80-100 | |
| 0 | 30 | 15 | 4 | 60-80 | 80-100 | 30F p/w 3wks episodic confusion - MRI = LT hyperdense lesion, ? cav mal. |
| | | | | 80-100 | 80-100 | |
| | | | | 60-80 | 80-100 | |
| 0 | 41 | 8 | 13 | 0-20 | 60-80 | 39 found down, aphasic, plegic on L on waking -> MIR = R ICA |
| | | | | 0-20 | 20-40 | |
| | | | | 20-40 | 0-20 | dissection + R hem stroke |
| 0 | 25 | 15 | 0 | 80-100 | 60-80 | 24F healthy p/w 5df fevers, HA, 2 GTCs, waxing and waning confusion -> viral encephalitis? |
| | | | | 80-100 | 80-100 | |
| | | | | 60-80 | 60-80 | |
| 0 | 38 | 15 | 2 | 20-40 | 80-100 | 38F w/ epilepsy p/w WHOL -> diffuse SAH (ACOM), neuro exam at baseline. |
| | | | | 20-40 | 60-80 | |
| | | | | 20-40 | 60-80 | |
| 1 | 48 | 15 | 4 | 40-60 | 80-100 | 48M p/w HA, new R temporal enhancing lesion -> resected -> AMS, suspected szs. |
| | | | | 60-80 | 60-80 | |
| | | | | 50-70 | 80-100 | |

**Neurologists' notes:** All the patients in this cohort are younger individuals with an exception of one 65y old female. The patients have largely similar prognosis as well as mix of chronic diseases. Most of them have high probability of a good outcome. Overall, matching seems reasonable w.r.t having similar 3 month prognoses based on admission data

**Matched Group 3**
Diameter: 43.34

| | | | | | | |
|---|---|---|---|---|---|---|
| 1 | 87 | 15 | 7 | 80-100 | 60-80 | 86M w/ 1 prior sz, p/w intermittent visual hallucinations c/f simple partial szs. CT = RT mass, c/w cav mal |
| | | | | 60-80 | 60-80 | |
| | | | | 20-40 | 60-80 | |
| 1 | 73 | 11 | 14 | 60-80 | 40-60 | 72M h/o anxiety on CLZ, AFib on dabigatran, p/w several GTC and confusion. |
| | | | | 60-80 | 80-100 | |
| | | | | 50-70 | 50-70 | |
| 0 | 72 | 15 | 11 | 80-100 | 40-60 | 71F w/ recent RF meningioma resection & recent wound infections p/w recurrent wound leakage & partial szs |
| | | | | 60-80 | 60-80 | |
| | | | | 70-90 | 40-60 | |
| 0 | 59 | 15 | 21 | 40-60 | 40-60 | 58F w/ complex history including |
| | | | | 0-20 | 40-60 | prior SAH and lupus cerebritis, p/w sepsis h/o and AMS |
| | | | | 20-40 | 40-60 | |
| 1 | 65 | 15 | 9 | 40-60 | 40-60 | 64 w/ CAD had nuclear stress test and persistent AMS and ?aphasia following procedure |
| | | | | 0-20 | 80-100 | |
| | | | | 20-40 | 60-80 | |
| 1 | 50 | 15 | 4 | 60-80 | 60-80 | 49M w/ suspected mitochondrial disease) p/w episodes of tunnel vision & unresponsiveness |
| | | | | 40-60 | 60-80 | |
| | | | | 40-60 | 40-60 | |
| 0 | 71 | 14 | 10 | 40-60 | 40-60 | 70F p/w progressive AMS x weeks -> lung CA, widespread mets including p-foss mass |
| | | | | 0-20 | 20-40 | |
| | | | | 20-40 | 30-50 | |
| 0 | 62 | 15 | 3 | 80-100 | 60-80 | 62F w/ R FT meningioma (resected yrs ago) p/w focal motor sz and L VFC and L sided neglect. |
| | | | | 40-60 | 60-80 | |
| | | | | 70-90 | 70-90 | |
| 1 | 73 | 13 | 9 | 60-80 | 40-60 | 73yM w/ NSCLC+brain mets focal szs, p/w acute AMS & increased tumor-related edema on NCHCT. |
| | | | | 40-60 | 20-40 | |
| | | | | 60-80 | 40-60 | |
| 1 | 74 | 14 | 13 | 80-100 | 40-60 | 73M w/ GBP, p/w 3wk decline in gait + confusion & episodes of sudden arm extension |
| | | | | 20-40 | 20-40 | |
| | | | | 50-70 | 30-50 | |
| 1 | 80 | 14 | 10 | 80-100 | 40-60 | 79M p/w 1d HA headache, confusion -> gtc x2, CSF: inc WBC; MRI = sphenoid sinusitis |
| | | | | 60-80 | 40-60 | |
| | | | | 80-100 | 30-50 | |

> **Neurologists' notes:** Most of the patients in this matched group are relatively older with higher APACHE II score and similar mix of hronic diseases. The cohort also has similar prognosis and high GCS scores. One caveat is the relatively large range of probability of high EA burden. Overall, this group seems to have more variablility in terms of disease severity and 3 month prognosis, compared to the other groups.

# Supplementary Material for Effects of Epileptiform Activity on Discharge Outcome in Critically Ill Patients: A Retrospective Cross-Sectional Study

**Table of Contents**

# Notations

**Table S1. Table of primary notations and corresponding definitions.**

| $C_i$ | Pre-admission covariates for patient i |
|---|---|
| $N_{i,j}$ | Hill's Coefficient for drug j in patient i |
| $ED50_{i,j}$ | ED50 for drug j in patient i |
| $W_{i,t,j}$ | Drug j administered in patient i at time t |
| $\overline{W}_{i,j}$ | $\sum_{t=1}^{T} W_{i,t,j}$ |
| $D_{i,t,j}$ | Drug concentration for drug j in patient i at time t |
| $Y_i$ | Post-discharge outcome for unit i |
| $Z_{i,t}$ | EA burden at time t for unit i |
| $E_{max,i}$ | Maximum EA burden in a 6 hour window for unit i |
| $E_{mean,i}$ | Average EA burden for unit i |

**Dataset characteristics. Note that all patients were ICU patients.**

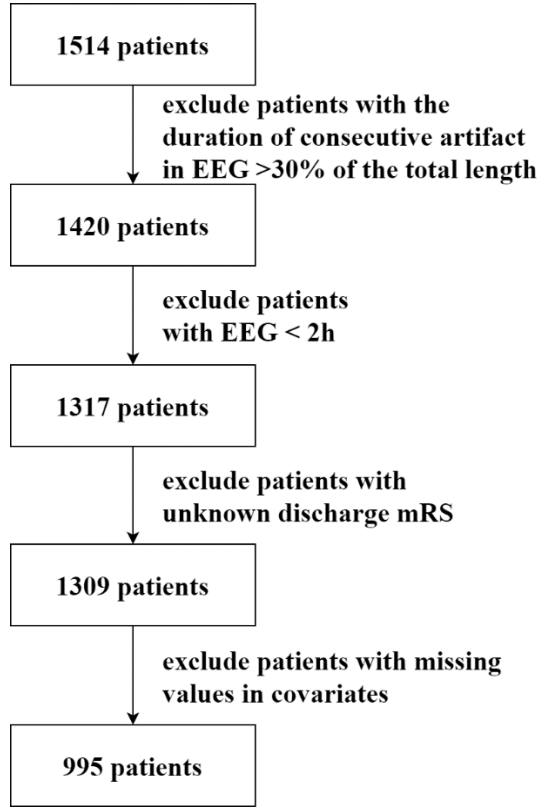

**Figure S1. Data flowchart showing the steps how the cohort was arrived at.**

**Table S2. Table describing full cohort characteristics across all demographics and medically relevant features. Median values are accompanied by interquartile range (IQR) and count values are accompanied by percentage (rounded off to nearest integer).**

| Variable | Value |
|---|---|
| Age, year, median (IQR) | 61 (48 -- 73) |
| Male gender, n (%) | 475 (48%) |
| Race | |
|     Asian, n (%) | 33 (3%) |
| Black / African American, n (%) | 72 (7%) |
| White / Caucasian, n (%) | 751 (75%) |
| Other, n (%) | 50 (5%) |
| Unavailable / Declined, n (%) | 84 (8%) |
| Married, n (%) | 500 (50%) |
| Premorbid mRS before admission, median (IQR) | 0 (0 -- 3) |
| APACHE II in first 24h, median (IQR) | 19 (11 -- 25) |
| Initial GCS, median (IQR) | 11 (6 -- 15) |
| Initial GCS is with intubation, n (%) | 415 (41%) |
| Worst GCS in first 24h, median (IQR) | 8 (3 -- 14) |
| Worst GCS in first 24h is with intubation, n (%) | 511 (51%) |
| Admitted due to surgery, n (%) | 168 (17%) |
| Cardiac arrest at admission, n (%) | 79 (8%) |
| Seizure at presentation, n (%) | 228 (23%) |
| Acute SDH at admission, n (%) | 146 (15%) |
| Take anti-epileptic drugs outside hospital, n (%) | 123 (12%) |
| Highest heart rate in first 24h, /min, median (IQR) | 92 (80 -- 107) |
| Lowest heart rate in first 24h, /min, median (IQR) | 71 (60 -- 84) |
| Highest systolic BP in first 24h, mmHg, median (IQR) | 153 (136 -- 176) |
| Lowest systolic BP in first 24h, mmHg, median (IQR) | 116 (100 -- 134) |
| Highest diastolic BP in first 24h, mmHg, median (IQR) | 84 (72 -- 95) |
| Lowest diastolic BP in first 24h, mmHg, median (IQR) | 61 (54 -- 72) |
| Mechanical ventilation on the first day of EEG, n (%) | 572 (57%) |
| Systolic BP on the first day of EEG, mmHg, median (IQR) | 148 (130 -- 170) |
| GCS on the first day of EEG, median (IQR) | 8 (5 -- 13) |
| History | |
|     Stroke, n (%) | 192 (19%) |
|     Hypertension, n (%) | 525 (53%) |
|     Seizure or epilepsy, n (%) | 182 (18%) |
|     Brain surgery, n (%) | 109 (11%) |
|     Chronic kidney disorder, n (%) | 112 (11%) |
|     Coronary artery disease and myocardial infarction, n (%) | 160 (16%) |
|     Congestive heart failure, n (%) | 90 (9%) |
|     Diabetes mellitus, n (%) | 201 (20%) |
|     Hypersensitivity lung disease, n (%) | 296 (30%) |
|     Peptic ulcer disease, n (%) | 50 (5%) |
|     Liver failure, n (%) | 46 (4%) |
|     Smoking, n (%) | 461 (46%) |
|     Alcohol abuse, n (%) | 231 (23%) |
|     Substance abuse, n (%) | 119 (12%) |
|     Cancer (except central nervous system), n (%) | 180 (18%) |
|     Central nervous system cancer, n (%) | 85 (8%) |
|     Peripheral vascular disease, n (%) | 41 (4%) |
|     Dementia, n (%) | 45 (4%) |
|     Chronic obstructive pulmonary disease or asthma, n (%) | 139 (14%) |
| Leukemia or lymphoma, n (%) | 22 (2%) |
| AIDS, n (%) | 12 (1%) |
| Connective tissue disease, n (%) | 47 (5%) |
| Primary diagnosis | |
| Septic shock, n (%) | 131 (13%) |
| Ischemic stroke, n (%) | 85 (8%) |
| Hemorrhagic stroke, n (%) | 163 (16%) |
| Subarachnoid hemorrhage (SAH), n (%) | 188 (19%) |
| Subdural hematoma (SDH), n (%) | 94 (9%) |
| SDH or other traumatic brain injury including SAH, n (%) | 52 (5%) |
| Traumatic brain injury including SAH, n (%) | 21 (2%) |
| Seizure/status epilepticus, n (%) | 258 (26%) |
| Brain tumor, n (%) | 113 (11%) |
| CNS infection, n (%) | 64 (6%) |
| Ischemic encephalopathy or Anoxic brain injury, n (%) | 72 (7%) |
| Toxic metabolic encephalopathy, n (%) | 104 (10%) |
| Primary psychiatric disorder, n (%) | 35 (3%) |
| Structural-degenerative diseases, n (%) | 35 (3%) |
| Spell, n (%) | 5 (1%) |
| Respiratory disorders, n (%) | 304 (31%) |
| Cardiovascular disorders, n (%) | 153 (15%) |
| Kidney failure, n (%) | 65 (7%) |
| Liver disorder, n (%) | 30 (3%) |
| Gastrointestinal disorder, n (%) | 18 (1%) |
| Genitourinary disorder, n (%) | 34 (3%) |
| Endocrine emergency, n (%) | 28 (3%) |
| Non-head trauma, n (%) | 13 (1%) |
| Malignancy, n (%) | 65 (7%) |
| Primary hematological disorder, n (%) | 24 (2%) |

**Anti-seizure medications**

Six drugs were studied: propofol, midazolam, levetiracetam, lacosamide, phenobarbital, and valproate. Propofol and midazolam are sedative antiepileptic drugs (SAEDs) which are given as continuous infusion, while the others are non-sedative antiepileptic drugs (NSAEDs) which are given as bolus. Only the period when there is EEG recording is used. The dose is normalized by body weight (kg). We use the half-lives from the literature (Table S2) for calculating the drug concentrations $D_{i,t,j}$ in the blood using the PK model.

**Table S3. Half-life for the anti-seizure medications used in the pharmacological modeling**

| **Drug generic name** | **Half-life** |
|---|---|
| Propofol | 20 minutes |
| Midazolam | 2.5 hours |
| Levetiracetam | 8 hours |
| Lacosamide | 11 hours |
| Phenobarbital | 79 hours |
| Valproate | 16 hours |
| Pentobarbital | 20 hours |
| Lorazepam | 15 hours |
| Diazepam | 43 hour |
| Fosphenytoin | 15 minutes |

**Binning of EA burden**

For statistical efficiency and interpretability, we bin the EA burden into 4 levels – mild, moderate, severe, very severe, as in Tables S5 and S6. Tables S4 and S6 discuss the correlation across various EA burden summaries.

**Table S4. Correlation of max, mean and median EA burden**

| Correlation Coefficient | $E_{max}$ | $E_{mean}$ | $E_{median}$ |
|---|---|---|---|
| $E_{max}$ | 1.00 | 0.71 | 0.70 |
| $E_{mean}$ | 0.71 | 1.00 | 0.99 |
| $E_{median}$ | 0.70 | 0.99 | 1.00 |

**Table S5. Binning of EA burden into 4 levels**

| EA Burden | Mild | Moderate | Severe | Very Severe |
|---|---|---|---|---|
| $E_{max}$ | 0 to 25% | 25% to 50% | 50% to 75% | 75% to 100% |
| Number of patients based on $E_{max}$ | 307 | 130 | 107 | 451 |
| $E_{mean}$ | 0 to 2% | 2% to 10% | 10% to 30% | 30% to 100% |
| Number of patients based on $E_{mean}$ | 284 | 232 | 220 | 259 |

Table S6. Correlation of max and mean EA burden discretized into 4 levels

| $E_{mean}$ \ $E_{max}$ | [0,0.25] | [0.25,0.5] | [0.5,0.75] | [0.75,1.0] | Grand Total |
|---|---|---|---|---|---|
| [0,0.02] | 263 | 20 | 1 | | 284 |
| [0.02,0.1] | 44 | 92 | 54 | 42 | 232 |
| [0.1,0.3] | | 18 | 49 | 153 | 220 |
| [0.3,1.0] | | | 3 | 256 | 259 |
| **Grand Total** | **307** | **130** | **107** | **451** | **995** |

**Drug Use Pattern**

Table S7. Number of patients with EA and acute system illness who are receiving non-sedating AEDs.

| | |
|---|---|
| Total # of patients | 995 |
| # of patients with EA | 960 |
| # of patients with EA and sepsis/shock and received NSAED | 79 (8%) |
| # of patients with EA and cardiovascular disorder and received NSAED | 83 (8%) |

| # of patients with EA and renal failure and received NSAED | 32 (3%) |
| # of patients with EA and liver disorder and received NSAED | 21 (2%) |
| # of patients with EA and endocrine emergency and received NSAED | 16 (2%) |
| # of patients with EA and non-head trauma and received NSAED | 12 (1%) |

Table S8. Co-administration of ASMs in patients.

|  | lacosamide | levetiracetam | midazolam | pentobarbital | phenobarbital | propofol | valproate | lorazepam | diazepam | fosphenytoin |
|---|---|---|---|---|---|---|---|---|---|---|
| **lacosamide** | 76 | 66 | 20 | 2 | 4 | 38 | 10 | 26 | 2 | 30 |
| **levetiracetam** | 66 | 620 | 79 | 3 | 16 | 353 | 30 | 112 | 9 | 89 |
| **midazolam** | 20 | 79 | 99 | 4 | 6 | 82 | 13 | 18 | 2 | 29 |
| **pentobarbital** | 2 | 3 | 4 | 5 | 0 | 4 | 1 | 0 | 0 | 3 |
| **phenobarbital** | 4 | 16 | 6 | 0 | 24 | 16 | 1 | 4 | 2 | 4 |
| **propofol** | 38 | 353 | 82 | 4 | 16 | 479 | 20 | 60 | 9 | 78 |
| **valproate** | 10 | 30 | 13 | 1 | 1 | 20 | 41 | 16 | 1 | 5 |
| lorazepam | 26 | 112 | 18 | 0 | 4 | 60 | 16 | 148 | 2 | 37 |
| diazepam | 2 | 9 | 2 | 0 | 2 | 9 | 1 | 2 | 14 | 1 |
| fosphenytoin | 30 | 89 | 29 | 3 | 4 | 78 | 5 | 37 | 1 | 113 |

**A framework for interpretable causal inference**

Here we describe the causal inference method used to estimate the potential outcomes. Given the stakes involved and the high level of noise in the data, we chose an interpretable causal inference method, MALTS, to estimate cause-effect relationships. MALTS is a matching method that learns a distance metric using a subset of data as training set. Further, the learned metric is used to produce high-quality matched groups on the rest of the units (also called as estimation set). These matched groups are used to estimate heterogeneous causal effects with high accuracy. Previous work on MALTS shows that it performs on-par with contemporary black-box causal machine learning methods while also ensuring interpretability[1].

The objective function of MALTS was designed to estimate the contrast of potential outcomes under binary "treatment". Here, we adapt it to estimate conditional average potential outcomes for n-ary "treatment." For our problem, there are 4 × 2 "treatment" arms – four levels of EA burden crossed by whether the patients were treated with ASMs. We construct the matched group $G_i$ for each patient i by matching on $X_i = \{C_{i,\cdot}, N_{i,\cdot}, ED_{50,i,\cdot}\}$, which is the vector of pre-admission covariates and PD parameters. For each matched group, MALTS also return diameter – the distance between the query patient and the farthest matched patient. We dropped/pruned all matched groups whose diameter were more $d_{prune}$. Here, we chose $d_{prune}$ equal to p, the number of covariates. As we standardize each covariate before matching, the variance across each covariate is equal to 1. Thus, intuitively, pruning matched groups with diameter more than p allows pruning units that are outliers in across all covariates. Next, we estimate $Pr\left[Y(E_{max} = e, \overline{W} = \delta) = 1 | X = X_i\right]$ by averaging the observed outcomes for units in the matched group $G_i$ with $E_{max}$ equals e and $\overline{W}$ equals δ: $\widehat{Pr}[Y(E_{max} = e, \overline{W} = \delta) = 1 | X = X_i] = \frac{\sum_{j \in G_i} 1[E_{max,j}=e, \overline{W}_j=\delta] Y_i}{\sum_{j \in G_i} 1[E_{max,j}=e, \overline{W}_j=\delta]}$.

We use an analogous estimator for $E_{mean}$.

MALTS' estimates of the conditional average potential outcome are interpretable because it is computed with the units in the matched groups. These matched groups can be investigated by looking at the raw data to examine their cohesiveness. One can immediately see anything that may need troubleshooting and determine how to troubleshoot it. For instance, if the matched group does not look cohesive, the learned distance metric might need troubleshooting. Or, processing of the EEG signal might need troubleshooting if the $EA_{max}$ values do not appear to be correct. Or, the PK/PD parameters might need troubleshooting if patients who appear to be reacting to drugs quickly are matched with others whose drug absorption rates appear to be slower, when at the same time, the PK/PD parameters appear similar.

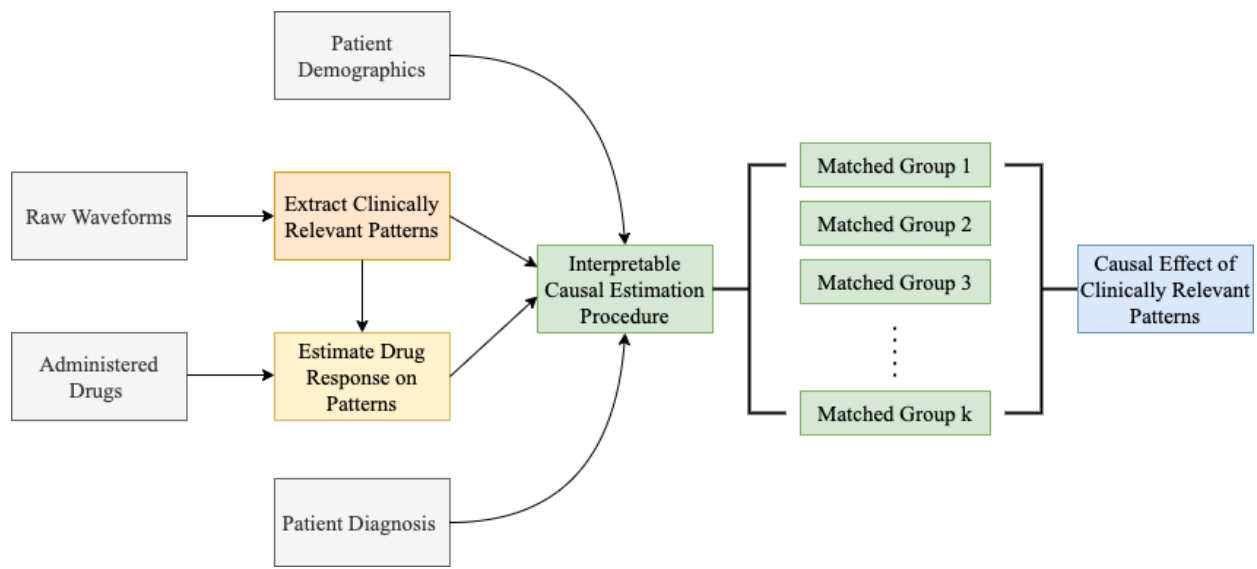

**Figure S2. The overall analysis framework. It consists of four parts (indicated by different colors): EA burden computation (orange block), individual PK/PD modeling (yellow block), MALTS matching (green blocks), and effect estimation (blue block).**

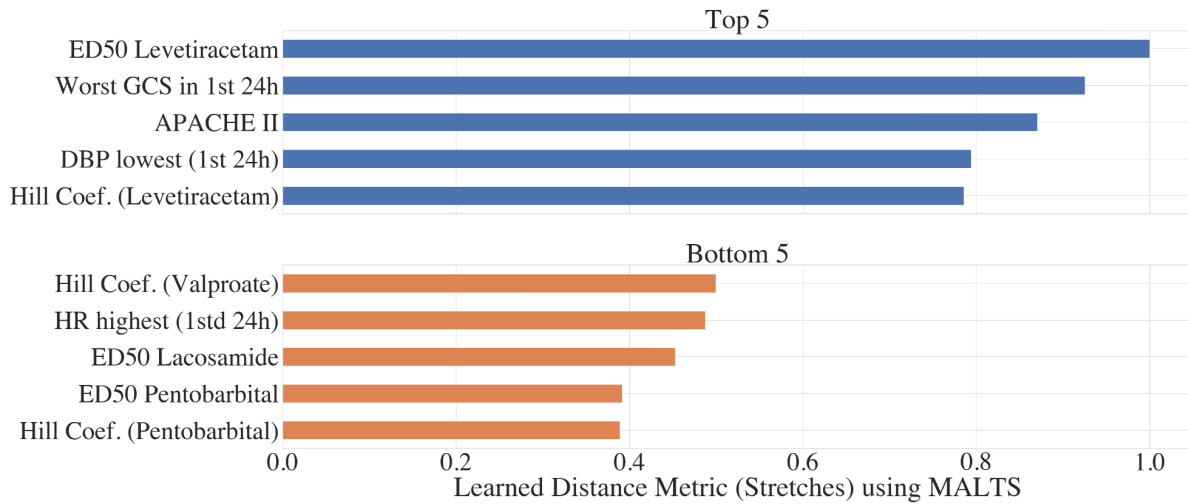

**Figure S3.** The plot shows variables that are the most important (top) and least important (bottom) to match on, as computed by MALTS, for studying the effect of the maximum EA burden $E_{max}$. BP = blood pressure; Coef. = coefficient.; ED50 = concentration of the drug that reduces EA burden by 50%; Hx = History; GCS = Glasgow Comma Score.

**Pharmacological modeling**

Doctors dynamically modify the type and dosage of ASM using the current EA observation, previous treatment, and patient's responsiveness to these treatments. This cyclical relationship potentially confounds the relationship between EA and a patient's final outcome. The heterogeneity in a patient's responsiveness to ASMs can be due to a variety of factors such as past medical history, current medical conditions, age, etc. However, the infrequency of some rare medical conditions makes it difficult to learn a nonparametric model of drug response that incorporates all relevant medical factors. To account for this, we leveraged the domain knowledge from pharmacology and use a one-compartment pharmacokinetic/pharmacodynamic (PK/PD) mechanistic model to estimate drug response as a function of ASM dose. The parameters of the PK/PD model can be interpreted as high-dimensional propensity scores that summarize a patient's responsiveness to a drug regime, such that any two patients with similar PK/PD parameters will exhibit similar responses under identical drug regimes.

We used a single-compartment PK model to estimate the blood concentration $D_{i,t,j}$ of ASM j in patient i at time t, and Hill's PD model[2] to estimate a short-term response to drugs:

$$\frac{dD_{i,t,j}}{dt} = -\frac{1}{\kappa_j} D_{i,t,j} + W_{i,t,j}, \tag{1}$$

$$Z_{i,t} = 1 - \sum_j \frac{D_{i,t,j}^{N_{i,j}}}{D_{i,t,j}^{N_{i,j}} + ED_{50,i,j}^{N_{i,j}}}. \tag{2}$$

Here $\kappa_j$ is the average half-life of the drug (see Table S3 for half-lives), $W_{i,t,j}$ is the body weight-normalized drug administration rate in units of mg/kg/h, $N_{i,j}$ represents how responsive the patient is to drug j, and $ED_{50,i,j}$ is the dosage required to reduce the patient's EA burden by 50%. Since $N_{i,j}$ (the Hill coefficient) is constrained to be non-negative, a positive correlation between drug concentration and EA burden results in an $N_{i,j}$ value of 0. The PD parameters were fit using least square method. The estimated PD parameters reflect wide heterogeneity across patients.

**EEG pre-processing and artifact detection**

The raw EEG signals were notch filtered at 60Hz, high-pass filtered at 0.5Hz, and resampled to 128 Hz. The signals were then re-referenced into 16-channel double banana bipolar montage[3,4] containing FP1-F7, F7-T3, T3-T5, T5-O1, FP2-F8, F8-T4, T4-T6, T6-O2, FP1-F3, F3-C3, C3-P3, P3-O1, FP2-F4, F4-C4, C4-P4, and P4-O2.

We define an artifact as a 10-second window in which one of the two following criteria is met: (1) total power is lower than first quartile minus 3× interquartile range (IQR) or higher than third quartile plus 3× IQR across total power of all 10-second windows; (2) the spectrum does not follow the 1/f rule, i.e., the slope of log-power spectral density is higher than the third quartile plus 3× IQR across the slops of all 10-second windows. We evaluated whether each 10 second window was an artifact.

**Human and deep learning EA labeling**

The human labels for training the deep learning model were derived as follows. First, 82 randomly selected patients' spectrograms were computed using 10-second windows. Neurologists used a labeling assistance tool NeuroBrowser[3] to label each 10-second EEG segment into one of six patterns: seizure (Sz), lateralized periodic discharges (LPD), generalized periodic discharges (GPD), lateralized rhythmic delta activity (LRDA), generalized rhythmic delta activity (GRDA), and normal. Labels of the 82 patients were used as the seed in deep active learning to train a deep neural network to automatically classify these patterns[4].

The deep neural network is a DenseNet with 7 blocks (Figure S5). Each block included 4 dense layers. Each dense layer is comprised of 2 convolutional layers and 2 exponential linear unit (ELU) activations. In between each dense block was a transition block consisting of an ELU activation, a convolutional layer, and an average pooling layer. There were 6 transition blocks in total. The last two layers of DenseNet were a fully connected layer followed by a softmax layer. The loss function includes Kullback-Leibler divergence inversely weighted by the class ratio to account for imbalance among the EA classes. After fitting, it was observed that DenseNet's classifications were much more volatile than the original data, with predictions abruptly changing from normal brain activity to EA patterns. This highlighted a limitation of traditional EEG classification from images, as the images were fed independently with no context about neighboring images beyond the 10-second window given. To correct for this volatility, the results of DenseNet were smoothed using a Hidden Markov Model. To smooth to a similar degree as the human labeled data, the probabilities of the transition matrix were fit on the 82 human-labeled patients. These probabilities were then used as the hidden state to smooth the output from DenseNet. We made the HMM first order due to precedent of first order HMMs providing good smoothing for other EEG problems[5].

The results of the automatic EA annotator resulted in accuracy for Seizure at 39% (human inter-rater agreement 42%), GPD at 62% (62%), LPD at 53% (58%), LRDA at 38% (38%), GRDA at 61% (40%), and others/artifact at 69% (75%), therefore matching human performance. We further combined the classification into binary classes, EA (seizure/GPD/LPD/LRDA) vs: non-EA (GRDA/normal) (Figure S6) to reduce the chance of error since these patterns are intrinsically on a continuous spectrum.

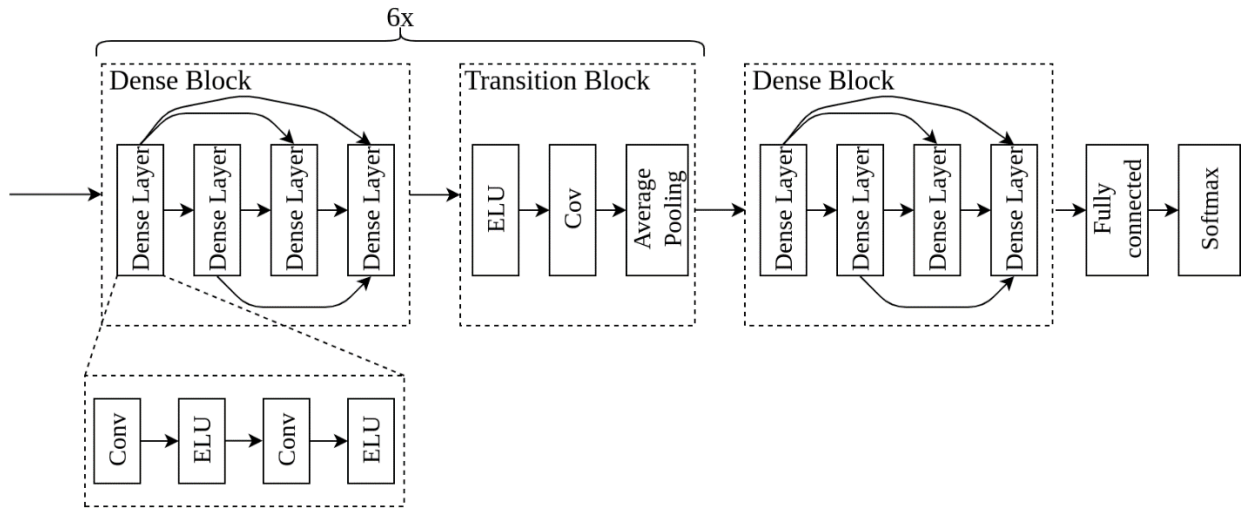

**Figure S5. Structure of the DenseNet for automatic EA labeling.**

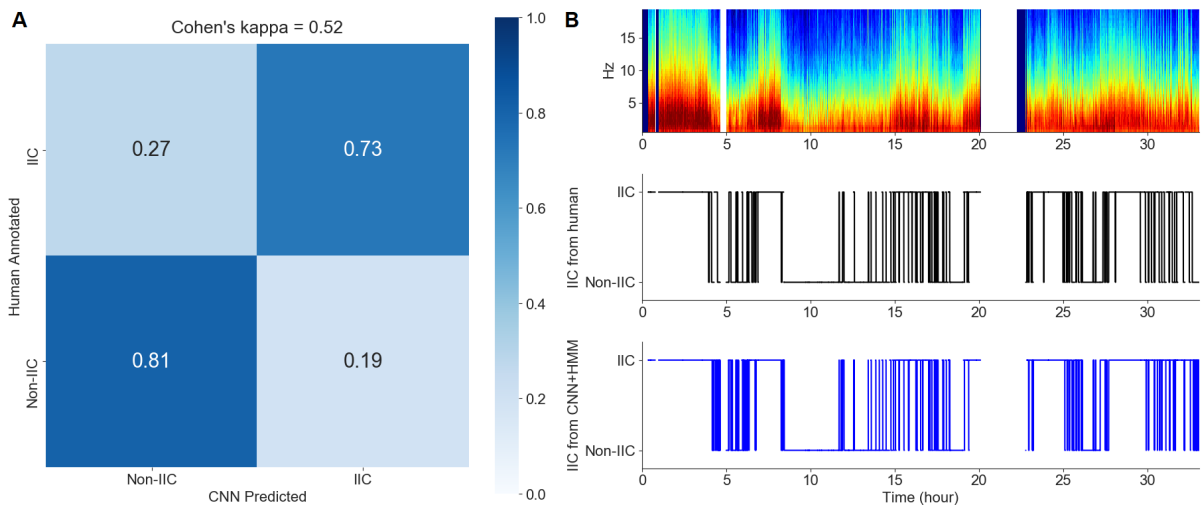

Figure S6. Performance and example of CNN predicted EA labels. (A) Confusion matrix for the CNN prediction vs. human annotation, where each row represents the fraction of 2-second segments classified into EA (seizure/GPD/LPD/LRDA) or Non-EA (GRDA/normal). The overall Cohen's kappa is 0.52. (B) The top panel shows the spectrogram of the EEG signal of one example patient; the middle panel shows EA patterns annotated by a human expert for every 2 second interval. The bottom panel shows the EA pattern annotated by the CNN followed by HMM smoothing.

**Missingness pattern**

Seizure risk prediction methods such as 2HELPS2B or TERSE are often used in practice to estimate the risk of subsequent EA (specifically, seizures). Such methods often do not account for treatment effects. Thus, these methods do not provide any information about the effect of the ASM treatment. Thus, such prediction methods cannot be used to impute missing EEG. Hence, we exclude units with less than 2hours of EEG recording as that is not the population of our interest.

To check for possible selection bias due to inclusion/exclusion criteria shown in Figure S1, we compared the discharge mRS in patients with different missing conditions shown in Figure S7, where some of them were excluded in this cohort. We used the Mann-Whitney U test (nonparametric t-test) to compare the medians, since mRS does not follow a normal distribution. The results showed that the medians of discharge mRS in patients with EEG, versus that in patients without EEG, are not significantly different (p = 0.32); similarly, the medians in patients with both EEG and drug data, versus that in patients without EEG or drug data, were not significantly different neither (p = 0.39). Therefore, the missingness pattern can be considered as not influencing our results.

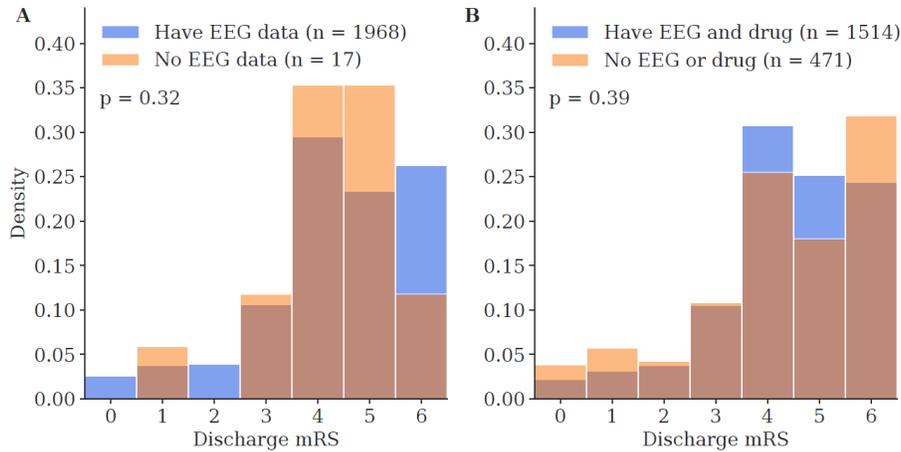

**Figure S7. Missingness pattern. (A) The histogram of patients' discharge mRS (possible values are 0,1,2,3,4,5,6). The two subsets that are compared are patients who have EEG data (n = 1968) vs. patients who do not have EEG data. To make the subsets comparison, the y-axis shows the density instead of the count. The p-value is from the Mann-Whitney U test of the two subsets. (B) Similar to A, but for patients who have EEG and drug data vs. patients who do not have EEG or drug data.**

**Sensitivity analysis**

Sensitivity analysis includes checking if the conclusion changes in terms of violation of assumptions, unobserved confounding, and measurement imprecision. The assumptions for valid causal inference include: 1) pre-admission covariates and PD parameters are both potential sources of confounding and thus need to be controlled for; and 2) the post-discharge outcome, Y, is directly affected by both the level of EA burden and the presence of anti-seizure medications W. Here, we demonstrate how the estimation of potential outcome can vary with these assumptions.

*Violation of assumption 1): The need to control for pre-admission covariates and PD parameters*

Previously, it was posited that pre-admission covariates such as age and diagnosis and PD parameters could be large sources of confounding in the estimation of average potential outcomes. In this section, we investigate this assumption by having MALTS create matched groups based on fewer and fewer factors and comparing the resulting average potential outcomes.

The right part of Figure S8 shows the estimated average potential outcome when MALTS controls for only age. The results do not show a monotonic relationship between EA burden and average potential outcome. Matching on all pre-admission covariates but no PD parameters using MALTS (middle part) also fails to yield a clear monotonic relationship between EA burden and post-discharge potential outcomes. Further, the uncertainty in the estimates is large, and the results tend to underestimate the probability that a patient would leave the hospital impaired or dead.

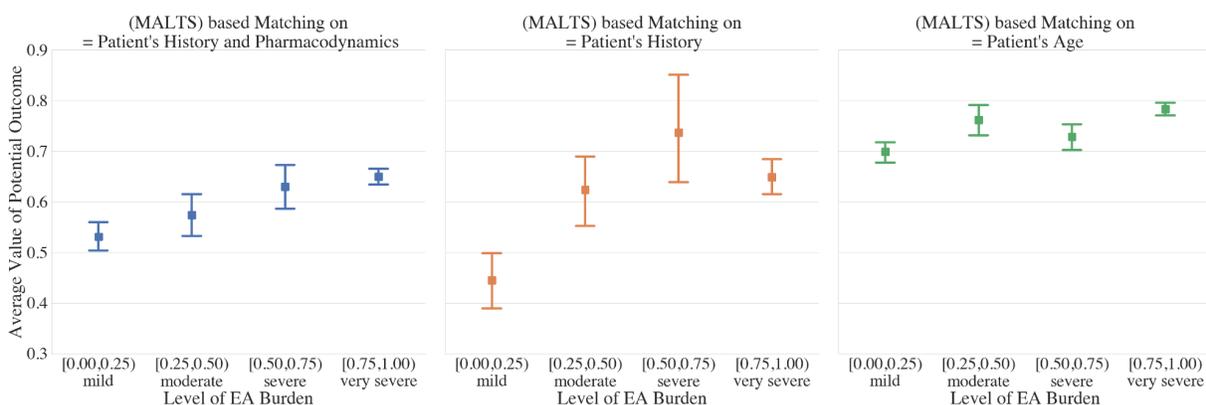

**Figure S8. Estimated average potential outcome for different $E_{max}$. This is by matching on (left) all pre-admission covariates and PD parameters, (middle) all pre-admission covariates, and (right) only age of the patients.**

*Violation of assumption 2): Post-discharge outcome are a function of the level of EA burden and the presence of anti-seizure medications*

Here we compare our approach to a naive average approach which posits that EA burden is the only causal factor, an outcome modeling approach that treats all factors in our study as having a direct causal effect on the outcome, and a propensity score matching approach, which performs a causal estimation under different assumptions.

In the naive average approach, at each EA burden (Figure S9, left), 1/3 of the data was left out and the probability of poor outcome was computed on the remaining 2/3 of the data. This procedure was repeated 15 times. The choice of 15 and 2/3 was done to match as closely as possible the 15 replicates and 2:1 training to testing ratio that was used by MALTS.

In the outcome modeling approach (Figure S9, middle), we performed a logistic regression where we regress the post-discharge outcome against EA burden, the presence of anti-seizure medications, and all of the factors that MALTS matched such as pre-admission covariates and PD parameters. Note that this approach assumes that there are no interactions between the regressors, which is contrary to our understanding of the treatment procedure, as factors such as age and diagnosis have a known interaction with a patient's response to anti-seizure medications. We also performed 15 replicates of a logistic regression with the same 2:1 train/test split.

In the propensity score matching approach, unlike MALTS which matches together patients directly on their covariates, propensity score matching is based on matching patients based on a their probability of being within the treatment or control arm. This makes the stronger assumption that the probability of being within the treatment or control arm can be modeled parametrically, in this case as using a logistic regression.

As shown in Figure S9, the results of these three approaches all yielded similar results. This differs from the original MALTS result in the top left of Figure S8 which shows a clear monotonic relationship between EA burden and average potential outcome.

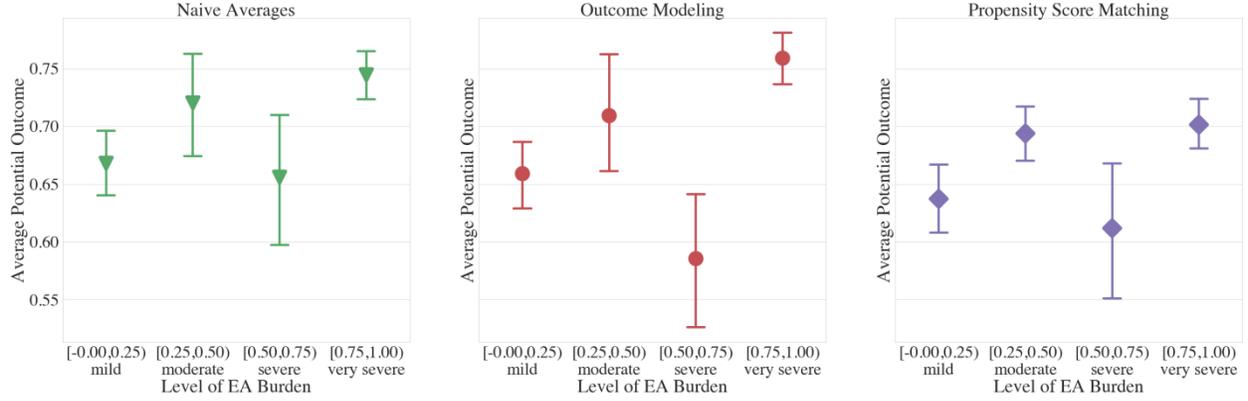

**Figure S9. Estimated average potential outcomes, computed using (left) the naïve average approach, (middle) the outcome modeling approach, and (right) the propensity score matching.**

*Unobserved confounding*

Here we study the sensitivity to an unobserved confounder that correlates patients' post-discharge outcome with $E_{max}$. We would like to see if the presence of an unobserved confounder we failed to control for could have biased our inferences. We can encode the effect of an unobserved confounder using a selection bias function q(e) with sensitivity parameter ψ, where e is EA burden. This approach is similar to the one proposed in Blackwell[6]. We parameterize q(·) as a logarithmic function of e.

$$q(e) = E[Y_i(e,0)|E_{max,i} = e, \underline{W_i} = 0] - E[Y_i(e,0)|E_{max,i} \neq e, \underline{W_i} = 0]$$

$$= \psi \ln \ln (e + 1)$$

When ψ is positive (negative), this indicates that patients with observed bad (good) outcomes also have high observed EA burden. This parametric form also assumes that a patient with low $E_{max}$ is affected less by an unobserved confounder U compared to a unit with higher $E_{max}$ with the marginal increase tapering off as the $E_{max}$ increases. This is congruent with the neurologist's intuition that individuals with normal brain activity will be affected less by an unobserved confounder U.

To perform the sensitivity analysis, we apply the following debiasing to the observed outcome and re-estimate the average potential outcomes:

$$Y_i^{debiased} = Y_i - q(E_{max,i})\left(1 - P(E_{max,i}|X = X_i)\right).$$

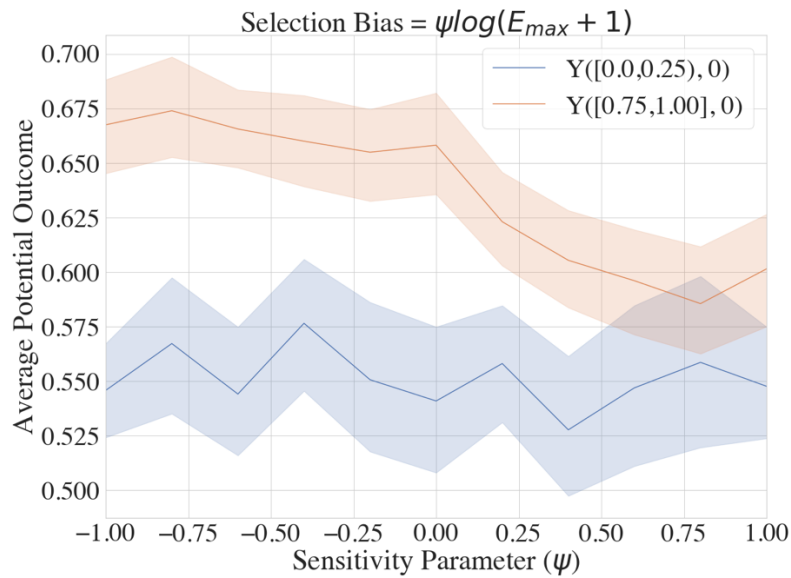

**Figure S10. Sensitivity to unobserved confounding**. **The results show that even at very high levels of unobserved confounding, the effect of EA burden is not lost, indicating a degree of robustness in our results.**

In Figure S10, the sensitivity analysis results showed that point estimate of potential outcome under very severe EA burden is always worse than the potential outcomes under mild EA burden for a range of sensitivity parameter ψ between [-1,1]. We further find that our inference is statistically significant for a wide range of ψ: -1.0 ≤ ψ ≤ 0.5. The sensitivity highlights that the conclusions from our study and analysis are not sensitive to high levels of unobserved confounding.

*Imprecision in EA burden*

Throughout the analysis, the summaries of EA burden, $E_{max}$ and $E_{mean}$ were quantized into four equally sized groups. This was done in accordance with clinician recommendations. Here we evaluate the sensitivity of the analysis to these decisions. Specifically, we consider $E_{max} \in \{[0, \rho_1), [\rho_1, 0.5), [0.5, \rho_2), [\rho_2, 1]\}$ where the analysis in the main text specifies $\rho_1 = 0.25$ and $\rho_2 = 0.75$. The interpretation of these parameters is as follows: the mild EA burden category allows for no more than $100 \times \rho_1$ percent of a six hour window to be spent with EA and the very severe EA burden category allows for no less than $100 \times \rho_2$ percent of a six hour window to be spent with EA. By varying these parameters, we redefine which individuals are considered mild versus very severe EA during the analysis.

As shown in Figure S11, we observe following:

- The potential outcome under mild EA burden, i.e., $E[Y([0, \rho_1), \underline{W}=0)]$, is mildly sensitive to changes in $\rho_2$ which is expected. Further, we observe that the gradient of the same with respect to $\rho_1$ is relatively flat, and $E[Y([0, \rho_1), \underline{W}=0)]$ is bounded between 0.53 and 0.6 for $\rho_1 \in [0.1; 0.4]$.

- Analogously the potential outcome under very severe EA burden, i.e., $E[Y([\rho_2, 1], \underline{W}=0)]$, is mildly sensitive to changes in $\rho_1$ and its gradient with respect to $\rho_2$ is relatively flat, and $E[Y([\rho_2, 1], \underline{W}=0)]$ is bounded between 0.645 and 0.705 for $\rho_2 \in [0.6, 0.9]$.

- The point estimates of $E[Y([0, \rho_1), \underline{W}=0)]$ are always strictly less than the point estimates of $E[Y([\rho_2, 1], \underline{W}=0)]$.

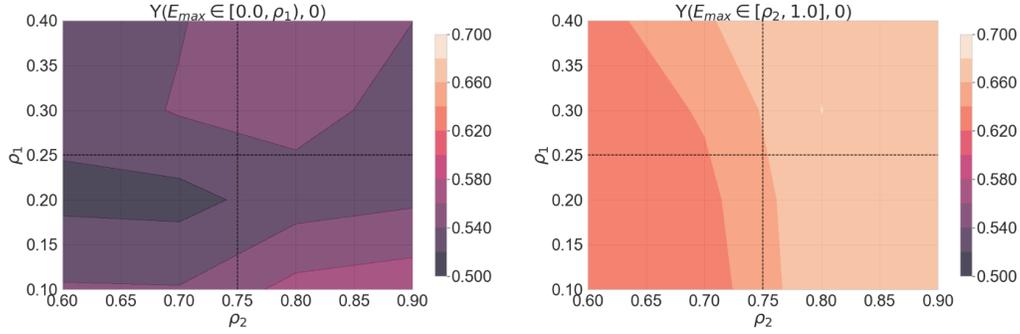

**Figure S11. Sensitivity to quantization of EA burden into four levels.** $\rho_1$ is the boundary between mild and moderate EA burden and $\rho_2$ is the boundary between severe and very severe EA burden. The contour plot shows estimated average potential outcomes, $E[Y([0, \rho_1), \underline{W}=0)]$ and $E[Y([\rho_2, 1], \underline{W}=0)]$, for a range of $\rho_1$ and $\rho_2$. We find that the estimates do not change by a large amount as $\rho_1$ and $\rho_2$ change as indicated by the numbers on the color bar, as compared to numbers in Figure 3 in the main text.

*Granular partitioning of EA*

Here, we study the average effect of $E_{max}$ if it is binned more granularly. Particularly, we are interested in checking if $E_{max} = 20\%$ is an inflection point. Figure S12 suggested conjecture of an inflection point at 20% might be reasonable. However, matching becomes more difficult and the results become more noisy and uncertain as the number of bins increases, thus for the main results we determined that our 4 levels provided a good balance between statistical estimability and resolution in EA levels.

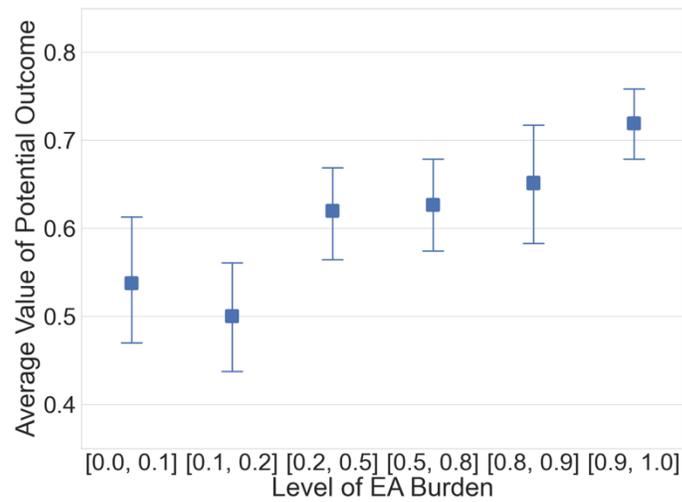

**Figure S12. Granular Binning of $E_{max}$ and its effect on potential mRS.** The EA max is binned into 6 levels instead of 4 such that the levels at the extremes are more granular. The trend of effect of EA remains similar to the one described in Figure 2.

**Distribution of Matched Group sizes**

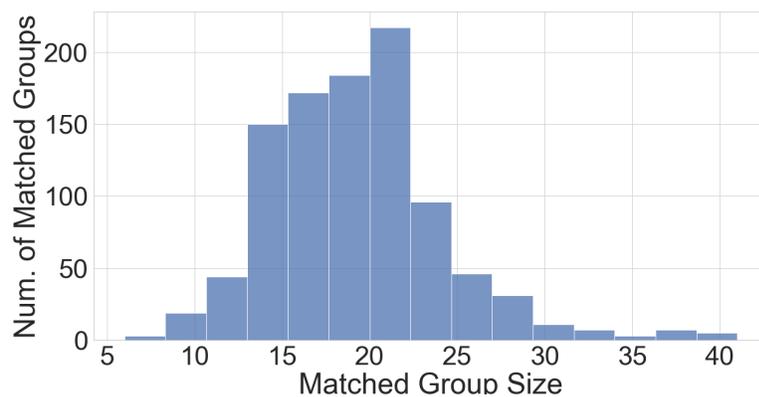

**Figure S13. Histogram of Matched Group sizes.** The distribution of MG size is a unimodal distribution with maximum size of 40 and minimum size of 6.

## Access to code repository

We provide the code base used for generating the results in the following GitHub repository: https://github.com/Hockey86/iic_causal_inference

## STROBE Statement

Checklist of items that should be included in reports of observational studies

|  | Item No. | Recommendation | Page No. | Relevant text from manuscript |
|---|---|---|---|---|
| **Title and abstract** | 1 | (*a*) Indicate the study's design with a commonly used term in the title or the abstract | 1 | Title: Effects of Epileptiform Activity on Discharge Outcome in Critically Ill Patients: A Retrospective Cross-Sectional Study |
|  |  | (*b*) Provide in the abstract an informative and balanced summary of what was done and what was found | 2 | In Summary |
| **Introduction** |  |  |  |  |
| Background/rationale | 2 | Explain the scientific background and rationale for the investigation being reported | 3 | In Introduction |
| Objectives | 3 | State specific objectives, including any prespecified hypotheses | 3 | "Our objective is to quantify the heterogeneous effects of EA with interpretability-centered approach where a physician can verify the quality of every single step in the analysis" |
| **Methods** |  |  |  |  |
| Study design | 4 | Present key elements of study design early in the paper | 3 | "This is a retrospective cross-sectional study" |
| Setting | 5 | Describe the setting, locations, and relevant dates, including periods of recruitment, exposure, follow-up, and data collection | 3 | "This is a retrospective cross-sectional study of patients admitted to the Massachusetts General Hospital (MGH) between September 2011 to February 2017" |
| Participants | 6 | (*a*) Cohort study—Give the eligibility criteria, and the | 3 | "The inclusion criteria are (1) a |
| | | sources and methods of selection of participants. Describe methods of follow-up<br>*Case-control study*—Give the eligibility criteria, and the sources and methods of case ascertainment and control selection. Give the rationale for the choice of cases and controls<br>*Cross-sectional study*—Give the eligibility criteria, and the sources and methods of selection of participants | | clinical neurophysiologist or epileptologist read the reports of EEG findings in the electronic health record of MGH and identified electrographic EA; and (2) at least 18 years old. The exclusion criteria (Figure S1) are (1) low EEG quality, where the duration of consecutive artifact (defined in "EEG pre-processing and artifact detection" in Supplementary Material) is more than 30% of the total length; (2) less than 2 hours of continuous EEG monitoring; and (3) missing outcome or covariates." |
| | | (*b*) *Cohort study*—For matched studies, give matching criteria and number of exposed and unexposed<br>*Case-control study*—For matched studies, give matching criteria and the number of controls per case | NA | NA |
| Variables | 7 | Clearly define all outcomes, exposures, predictors, potential confounders, and effect modifiers. Give diagnostic criteria, if applicable | 4 | In Methods, indicated by subsection titles |
| Data sources/ measurement | 8* | For each variable of interest, give sources of data and details of methods of assessment (measurement). Describe comparability of assessment methods if there is more than one group | 4 | In Methods, indicated by subsection titles |
| Bias | 9 | Describe any efforts to address potential sources of bias | 5 | In Methods, Statistical analysis, it includes confounding bias, selection bias, and sensitivity analysis |
| Study size | 10 | Explain how the study size was arrived at | Page 2 in Supplementary Material | Figure S1 |
| Quantitative variables | 11 | Explain how quantitative variables were handled in the analyses. If applicable, describe which groupings were chosen and why | 4,5 And pages 11 and 12 in Supplementary Material | In Methods, Statistical analysis "Continuous data is presented using the median and interquartile range. Categorical data is presented using number and percentage." Exposure is grouped as written in "Exposure: EA burden". Sensitivity of grouping were described in Figure S11. |
| Statistical methods | 12 | (*a*) Describe all statistical methods, including those used to control for confounding | 5 | In Methods, Effect estimation through matching |
| | | (*b*) Describe any methods used to examine subgroups and interactions | 5 | In Methods, Neurologist review of the matched groups |
| | | (*c*) Explain how missing data were addressed | 4 and pages 8 and 9 in Supplementary Material | In Methods, under exposure, outcome, and covariates subsections. The sensitivity analysis for missing data is in Figure S7. |
| | | (*d*) *Cohort study*—If applicable, explain how loss to follow-up was addressed *Case-control study*—If applicable, explain how matching of cases and controls was addressed *Cross-sectional study*—If applicable, describe analytical methods taking account of sampling strategy | NA | NA |
| | | (*e*) Describe any sensitivity analyses | Pages 9 to 12 in Supplementary Material | In Supplementary Material there is a section "Sensitivity analysis" |
| **Results** | | | | |
| Participants | 13* | (a) Report numbers of individuals at each stage of study—eg numbers potentially eligible, examined for eligibility, confirmed eligible, included in the study, completing follow-up, and analysed | Page 2 in Supplementary Material | Figure S1 |
| | | (b) Give reasons for non-participation at each stage | Page 2 in Supplementary Material | Figure S1 |

| | | | | | |
|---|---|---|---|---|---|
| | | | (c) Consider use of a flow diagram | Page 2 in Supplementary Material | Figure S1 |
| Descriptive data | | 14* | (a) Give characteristics of study participants (eg demographic, clinical, social) and information on exposures and potential confounders | Will be inserted into the main text | Table 1 |
| | | | (b) Indicate number of participants with missing data for each variable of interest | 4 | In Methods, under exposure, outcome, and covariates subsections. |
| | | | (c) *Cohort study*—Summarise follow-up time (eg, average and total amount) | NA | NA |
| Outcome data | | 15* | *Cohort study*—Report numbers of outcome events or summary measures over time | NA | NA |
| | | | *Case-control study*—Report numbers in each exposure category, or summary measures of exposure | NA | NA |
| | | | *Cross-sectional study*—Report numbers of outcome events or summary measures | Will be inserted into the main text | Table 1 |
| Main results | | 16 | (*a*) Give unadjusted estimates and, if applicable, confounder-adjusted estimates and their precision (eg, 95% confidence interval). Make clear which confounders were adjusted for and why they were included | 6 | In Results |
| | | | (*b*) Report category boundaries when continuous variables were categorized | Will be inserted into the main text | Figure 2 |
| | | | (*c*) If relevant, consider translating estimates of relative risk into absolute risk for a meaningful time period | 4,5 | Explained in Methods, "We dichotomized mRS into poor (mRS ≥ 4: moderately severe disability) and favorable (mRS ≤ 3: moderate disability) outcomes" and "Our estimand of interest is the probability of a poor outcome if the patient has EA burden (Emax or Emean) equal to a given level in the absence of treatment." |
| Other analyses | | 17 | Report other analyses done—eg analyses of subgroups and interactions, and sensitivity analyses | 6 | In Results, subsection: Heterogeneity in effects for maximum EA burden, Interpretable matched group analysis, |

| | | | | Matched groups are validated by neurologists' chart review, and sensitivity analysis in the Supplementary Material |
|---|---|---|---|---|
| **Discussion** | | | | |
| Key results | 18 | Summarise key results with reference to study objectives | 7 | In Discussion, Results in context. |
| Limitations | 19 | Discuss limitations of the study, taking into account sources of potential bias or imprecision. Discuss both direction and magnitude of any potential bias | 8 | In Discussion, Limitations, "there are still possibilities of unmeasured confounding, such as illness presented before admission that positively contributes to both having high EA burden and poor outcome, hence the effect is overestimated. On the other hand, since this is a hospital-based population, their baseline outcome is already biased to the poor side compared to a healthy population, which makes it hard for the discharge outcome to be even worse, hence the effect is underestimated." |
| Interpretation | 20 | Give a cautious overall interpretation of results considering objectives, limitations, multiplicity of analyses, results from similar studies, and other relevant evidence | 8 | At the end of Discussion, "In summary, we quantified the harm caused by EA in ICU patients. We confirmed that EA burden indeed worsens outcomes, and this effect is more severe in specific subgroups." |
| Generalisability | 21 | Discuss the generalisability (external validity) of the study results | 7 | In Discussion, Generalizability |
| **Other information** | | | | |
| Funding | 22 | Give the source of funding and the role of the funders for the present study and, if applicable, for the original study on which the present article is based | 5 | At the end of Methods, "The funding source is not involved in any part of this research." |

*Give information separately for cases and controls in case-control studies and, if applicable, for exposed and unexposed groups in cohort and cross-sectional studies.

**Note:** An Explanation and Elaboration article discusses each checklist item and gives methodological background and published examples of transparent reporting. The STROBE checklist is best used in conjunction with this article (freely available on the Web sites of PLoS Medicine at http://www.plosmedicine.org/, Annals of Internal Medicine at http://www.annals.org/, and Epidemiology at http://www.epidem.com/). Information on the STROBE Initiative is available at www.strobe-statement.org.